\documentclass[letterpaper, 10 pt, conference]{ieeeconf}  
\IEEEoverridecommandlockouts

\usepackage{amsmath,amssymb,mathtools,mathrsfs}
\usepackage{graphicx}
\usepackage{url,comment}
\usepackage[noadjust,compress]{cite}
\usepackage{marginnote,color,xcolor}

\newcommand{\hs}{\hspace{-2mm}}
\newcommand{\mcal}[1]{\mathcal{#1}}
\newcommand{\Exp}{\mathbb{E}}
\newcommand{\Prob}{\mathbb{P}}
\newcommand{\ol}[1]{\overline{#1}}
\newcommand{\ul}[1]{\underline{#1}}
\newcommand{\iTheta}{\mathit{\Theta}}
\newcommand{\tr}[1]{{\rm Tr}\left(#1\right)}
\newcommand{\rmx}{{\rm x}}
\newcommand{\rmy}{{\rm y}}
\newcommand{\rmw}{{\rm w}}
\newcommand{\rmv}{{\rm v}}
\newcommand{\rmA}{{\rm A}}
\newcommand{\rmC}{{\rm C}}
\newcommand{\rmQ}{{\rm Q}}
\newcommand{\rmR}{{\rm R}}
\newcommand{\rmP}{{\rm P}}
\newcommand{\rmX}{{\rm X}}
\newcommand{\Ph}{P_{\rm h}}
\newcommand{\pt}{p_{\rm h}}
\newcommand{\Pc}{P_{\rm c}}
\newcommand{\pc}{p_{\rm c}}
\newcommand{\Pfh}{\Ph^{\rm f}}
\newcommand{\Bf}{B^{\rm f}}
\newcommand{\Kf}{K^{\rm f}}
\newcommand{\Tf}{T^{\rm f}}
\newcommand{\sigmaf}{\sigma^{\rm f}}
\newcommand{\tg}{0}
\newcommand{\tb}{1}

\DeclareMathOperator*{\argmin}{arg\,min}

\newtheorem{lem}{Lemma}
\newtheorem{defin}{Definition}
\newtheorem{theorem}{Theorem}
\newtheorem{prop}{Proposition}
\newtheorem{assum}{Assumption}

\renewcommand{\QED}{\QEDopen}

\title{\LARGE \bf Structural Monotonicity in Transmission Scheduling\\ for Remote State Estimation with Hidden Channel Mode}

\author{Hampei Sasahara
\thanks{*This work was supported in part by JSPS KAKENHI Grant Number JP24K17296 and JST-ASPIRE Program Grant Number JPMJAP2402.}
\thanks{Hampei Sasahara is with the Department of Information Physics and
Computing, Graduate School of Information Science and Technology, The University of Tokyo, Tokyo, 113-8656 Japan
        {\tt\small hsasahara@g.ecc.u-tokyo.ac.jp}.}%
}

\begin{document}

\maketitle
\thispagestyle{empty}
\pagestyle{empty}

\begin{abstract}
This study treats transmission scheduling for remote state estimation over unreliable channels with a hidden mode.
A local Kalman estimator selects scheduling actions, such as power allocation and resource usage, and communicates with a remote estimator based on acknowledgement feedback, balancing estimation performance and communication cost.
The resulting problem is naturally formulated as a partially observable Markov decision process (POMDP).
In settings with observable channel modes, it is well known that monotonicity of the value function can be established via investigating order-preserving property of transition kernels.
In contrast, under partial observability, the transition kernels generally lack this property, which prevents the direct application of standard monotonicity arguments.
To overcome this difficulty, we introduce a novel technique, referred to as state-space folding, which induces transformed transition kernels recovering order preservation on the folded space.
This transformation enables a rigorous monotonicity analysis in the partially observable setting.
As a representative implication, we focus on an associated optimal stopping formulation and show that the resulting optimal scheduling policy admits a threshold structure.
\end{abstract}

\section{INTRODUCTION}

Networked control systems are now ubiquitous, making remote state estimation over unreliable communication channels a fundamental task~\cite{sinopoli2004kalman,shi2010kalman}.
Packet losses and communication delays degrade estimation performance, highlighting the importance of transmission scheduling, including transmission decisions and power allocation~\cite{wu2012event}.
Notably, this issue has recently gained renewed attention from a security perspective due to adversarial interference and denial-of-service attacks~\cite{zhang2015optimal,li2017sinr,ishii2022security}.

Within the framework of transmission scheduling, an important line of work has examined the structural properties, notably the monotonicity of value functions and policies for the underlying Markov decision processes (MDPs) identified in~\cite{leong2015optimality}.
The underlying intuition is that the system performance degrades as the holding time becomes longer or as the channel mode deteriorates, which consequently worsens the value function. 
These results have since been extended to settings with time-varying channel modes, multiple parallel processes, and related generalizations~\cite{qi2017optimal,wu2018optimal,wu2020learning,wei2023double,sun2024optimal}.
Such structural monotonicity not only provides insight into the nature of optimal policies but also significantly reduces computational complexity for policy computation and implementation~\cite{roy2021online,nakhleh2022deeptop,krishnamurthy2025partially}.

However, existing analyses typically assume full observability of the channel mode, which is often hidden and not directly observable in practice.
Although extensions to hidden channel modes have been explored, structural monotonicity has not been achieved for the resulting partially observable Markov decision processes (POMDPs).
A common approach to monotonicity analysis under partial observability relies on total positivity of order two (TP2) properties of the underlying transition kernels.
This approach, however, fails because the relevant state-transition probabilities have been shown to violate TP2 in the presence of holding-time dynamics~\cite{liu2022rollout,sun2025optimal}, posing a fundamental obstacle to extending monotonicity results.
Consequently, alternative methods that do not rely on monotonicity have been proposed, including rollout-based approximations~\cite{liu2022rollout} and two-stage approaches that explicitly estimate the channel mode prior to scheduling~\cite{sun2024optimal}.

This study establishes structural monotonicity even under partial observability by introducing a novel technique, referred to as \emph{state-space folding}. 
The key idea is to transform the transition kernels onto a folded state space, where the hidden order in the original kernel, which is lost due to the shape of its support under holding-time dynamics rather than any intrinsic property of the kernel itself, becomes explicit. 
This transformation restores TP2 properties and enables monotonicity analysis under partial observability.
In particular, the analysis reveals that the value function is monotonically increasing in both the holding time and the belief of the unfavorable channel mode, consistent with the results under full observation.
As a representative implication of this analysis, we focus on an associated optimal stopping formulation where the scheduler chooses between continuation and termination.
It is further shown that the resulting optimal scheduling policy admits a threshold structure, with the stopping action becoming optimal once the holding time and the belief exceed a critical level.

The remainder of this paper is organized as follows.
Section~\ref{sec:math_pre} reviews mathematical preliminaries.
Section~\ref{sec:prob} formulates the problem as a POMDP.
Section~\ref{sec:monotonicity_analysis} introduces state-space folding, proves TP2 recovery of transformed kernels, and derives the monotonicity and threshold results.
Section~\ref{sec:num} presents a numerical example that validates the theoretical findings, and Section~\ref{sec:conc} concludes the paper.

\emph{Notation:}
We denote
the set of nonnegative integers by $\mathbb{Z}_+,$
the trace of a matrix $M$ by $\tr{M},$
the spectral radius of a matrix $M$ by $\rho(M)$,
the maximum singular value of a matrix $M$ by $\|M\|_2$,
and
the positive and negative definiteness of a symmetric matrix $M$ by $M\succ0$ and $M\prec 0$, respectively.
Throughout this paper, the terms ``increasing'' and ``decreasing'' are used in the weak sense to mean non-decreasing and non-increasing.

\section{Mathematical Preliminaries}
\label{sec:math_pre}

We review the mathematical preliminaries required for the subsequent analysis.
Throughout this section, every domain is assumed to be a subset of $\mathbb{Z}$ for notational simplicity, although the results are not restricted to the discrete case. 

We begin by defining first-order stochastic dominance (FSD) order and monotone likelihood ratio (MLR) order for probability distributions.
\begin{defin}
    Let $p_1(x)$ and $p_2(x)$ be probability mass functions.
    Then $p_2$ dominates $p_1$
    in the MLR order (denoted by $p_1\leq_{\rm r}p_2$) if
    $p_1(x_1)p_2(x_2)-p_1(x_2)p_2(x_1)\geq 0$
    for any $x_1< x_2$.
\end{defin}
Note that for a binary set $\mcal{X}=\{0,1\}$ MLR dominance $p_1\leq_{\rm r}p_2$ is equivalent to $p_1(1)\leq p_2(1)$.
Moreover, MLR dominance implies ordering in expectation for all increasing functions~\cite[Theorem 10.4]{krishnamurthy2025partially}.
\begin{lem}\label{lem:MLR_dom_FSD}
    If the MLR dominance $p_1\leq_{\rm r}p_2$ holds, $\sum_{x}v(x)p_1(x)\leq\sum_{x}v(x)p_2(x)$ for any increasing function $v:\mcal{X}\to\mathbb{R}$.
\end{lem}
Subsequently, we define TP2 for nonnegative functions.
\begin{defin}
    A nonnegative function $f(x,y)$
    is said to be totally positive of order two (TP2) if
    $f(x_1,y_1)f(x_2,y_2)-f(x_1,y_2)f(x_2,y_1)\geq 0$
    for any $x_1< x_2$ and $y_1< y_2$.
\end{defin}
For stochastic kernels, its induced transformation preserves the MLR order of the input distribution, and the converse also holds~\cite[Theorem 11.10]{krishnamurthy2025partially}.
\begin{lem}\label{lem:TP2_transformation}
    Let $P(y|x)$ be a stochastic kernel and define $q_i(y)\coloneqq \sum_{x\in\mcal{X}}P(y|x)p_i(x)$ for $i=1,2$.
    Then $q_1\leq_{\rm r}q_2$ for any $p_1\leq_{\rm r}p_2$ iff $P$ is TP2.
\end{lem}

TP2 is preserved under marginalization~\cite[Prop.~3.4]{karlin1980classes}.
\begin{lem}\label{lem:TP2_preservation}
    Let $f_1(x,y)$ and $f_2(x,y)$ be TP2 functions.
    Then $f(x,z)\coloneqq \sum_{y}f_1(x,y)f_2(y,z)$ is TP2.
\end{lem}

Furthermore, TP2 is preserved under summation with positive weights.
\begin{lem}\label{lem:TP2_weighted_sum}
    Let $f_i(x,y)$ be TP2 functions.
    Then $f(x,y)\coloneqq \sum_{i}\alpha_i f_i(x,y)$ is TP2 for any $\alpha_i\geq 0$.
\end{lem}

An important application of the TP2 property arises in Bayesian inference~\cite[Theorems 11.5,~11.11]{krishnamurthy2025partially}.
\begin{lem}\label{lem:Bayes_order_preservation}
    Define the posterior distribution $p^+(x|y)\coloneqq K(y|x)p(x)/(\sum_{x' \in\mcal{X}}K(y|x')p(x'))$ with a stochastic kernel $K(y|x)$.
    Then $p^+_1(\cdot|y)\leq_{\rm r} p^+_2(\cdot|y)$ for $p_1\leq_{\rm r} p_2$.
    Further, if $K(y|x)$ is TP2, then $p^+(\cdot|y_1)\leq_{\rm r} p^+(\cdot|y_2)$ for $y_1\leq y_2$.
\end{lem}

Another related application is the Topkis' monotonicity theorem~\cite[Theorem 10.2]{krishnamurthy2025partially} under submodularity.
\begin{defin}
    A function $Q(x,a)$
    is said to be submodular if $Q(x_2,a_2)-Q(x_2,a_1)\leq Q(x_1,a_2)-Q(x_1,a_1)$ for any $x_1\leq x_2$ and $a_1\leq a_2$.
\end{defin}
\begin{lem}\label{lem:Topkis}
    If $Q(x,a)$ is submodular, the minimum of $\argmin_{a}Q(x,a)$ is incrasing.
\end{lem}

\section{PROBLEM FORMULATION}
\label{sec:prob}

\subsection{Remote State Estimation, Transmission Scheduling, Channel Model}

Following the standard setting in remote state estimation~\cite{sinopoli2004kalman,shi2010kalman}, we consider the discrete-time linear time-invariant system
$\rmx_{t+1} = \rmA\rmx_t + \rmw_t$ and $\rmy_t = \rmC\rmx_t + \rmv_t,$
where $\rmx_t\in\mathbb{R}^n$ is the system state, $\rmy_t\in\mathbb{R}^m$ is the measured output, and $\rmw_t$ and $\rmv_t$ are  independent and identically distributed (i.i.d.) Gaussian random noises with zero mean and covariance matrices $\rmQ\succeq0$ and $\rmR\succ 0,$ respectively.
As in standard Kalman filtering, we assume that the pair $(\rmA,\rmC)$ is observable and $(\rmA,\sqrt{\rmQ})$ is controllable.

\begin{figure}[t]
  \centering
  \includegraphics[width=0.98\linewidth]{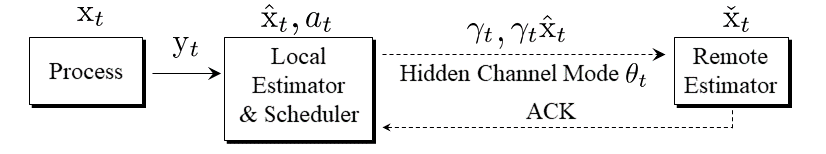}
  \caption{
  Overall system architecture.
  }
  \label{fig:sys_architecture}
\end{figure}

The overall system architecture considered in this study is illustrated by Fig.~\ref{fig:sys_architecture}.
First, the local estimator, often referred to as a smart sensor~\cite{li2017sinr}, estimates the state $\rmx_t$ at each time step according to the minimum mean-squared error (MMSE) criterion:
$\hat{\rmx}_t \coloneqq \Exp\left[ \rmx_t|\rmy_0,\ldots,\rmy_t \right],$
which can be computed using the standard Kalman filter~\cite{anderson2005optimal}.
The corresponding error covariance matrix
$\hat{\rmP}_t\coloneqq \Exp\left[ (\rmx_t-\hat{\rmx}_t)(\rmx_t-\hat{\rmx}_t)^{\sf T}| \rmy_0,\ldots,\rmy_t \right]$
satisfies the recursion
$\hat{\rmP}_{t+1} = \tilde{g}\circ h(\hat{\rmP}_t)$
where
$h(\rmX)\coloneqq \rmA\rmX\rmA^{\sf T}+\rmQ,$
$\tilde{g}(\rmX) \coloneqq \rmX-\rmX\rmC^{\sf T}(\rmC\rmX\rmC^{\sf T}+\rmR)^{-1}\rmC\rmX$~\cite{shi2010kalman}.
Under the controllability and observability assumption, there exists a steady-state error covariance $\bar{\rmP}$ such that $\tilde{g}\circ h(\bar{\rmP})=\bar{\rmP}$ for any initial condition~\cite{anderson2005optimal}.
In the following, for simplicity, we assume that the local estimation has already converged to the steady state, i.e., $\hat{\rmP}_t=\bar{\rmP}$ for any $t\in\mathbb{Z}_+$.

The local estimator transmits the estimated state $\hat{\rmx}_t$ to the remote estimator over an unreliable communication channel.
Let $\gamma_t=1$ and $\gamma_t=0$ indicate successful and failed transmissions at time $t$, respectively.
Denote by $\check{\rmx}_t$ the MMSE estimate of $\rmx_t$ at the remote estimator:
$\check{\rmx}_t \coloneqq \Exp\left[ \rmx_t|\gamma_0,\ldots,\gamma_t,\gamma_0\hat{\rmx}_0,\ldots,\gamma_t\hat{\rmx}_t \right].$
This remotely estimated state and the corresponding error covariance matrix, defined by
$\rmP_t\coloneqq\Exp[(\check{\rmx}_t-\rmx_t)(\check{\rmx}_t-\rmx_t)^{\sf T}|\gamma_0,\ldots,\gamma_t,\gamma_0\hat{\rmx}_0,\ldots,\gamma_t\hat{\rmx}_t],$
evolve recursively according to
\[
 (\check{\rmx}_{t+1},\rmP_{t+1}) = \left\{
 \begin{array}{ll}
  (\hat{\rmx}_{t+1},\bar{\rmP}), & {\rm if}\ \gamma_{t+1}=1,\\
  (\rmA\check{\rmx}_t,h(\rmP_t)), & {\rm otherwise}.
 \end{array}
 \right.\\
\]
The estimation performance is evaluated using the trace of the error covariance matrix, which depends solely on the holding time $\tau_t\in\mcal{T}\coloneqq\mathbb{Z}_+$, defined as the number of time steps since the most recent successful transmission.
The holding time evolves as
$\tau_{t+1}=0$ if $\gamma_{t+1}=1$ and $\tau_{t+1}=\tau_t+1$ otherwise.
Consequently, the estimation error at time $t$ can be expressed as $c_{\rm s}(\tau_t)\coloneqq\tr{\rmP_t}=\tr{h^{\tau_t}(\bar{\rmP})}$, where $h^\tau$ denotes the $\tau$-fold composition of $h$.
An important property is its monotonic dependence on the holding time~\cite[Lemma 4.1]{leong2015optimality}.
\begin{prop}\label{prop:monotonicity_cost}
    The estimation error $c_{\rm s}(\tau)$ is increasing with respect to the holding time $\tau\in\mcal{T}$.
\end{prop}

At every time step, the remote estimator sends an acknowledgment (ACK) signal to the local estimator indicating whether the most recent transmission was successful.
Based on the received ACK, the local estimator updates its information regarding the transmission outcome and the holding time.
We put the standard assumption that the ACK link is reliable~\cite{leong2015optimality,qi2017optimal}, reflecting the fact that the remote estimator is usually more powerful than the local sensor and that ACK packets carry only a very small amount of information, thereby requiring minimal communication capacity.


The scheduler implemented with the local estimator selects a transmission scheduling action $a_t\in\mcal{A}$, incurring an associated cost $c_{\rm a}(a_t)\in\mathbb{R}$, which influences the probability of successful transmission and the evolution of the channel mode, as described below.
Typical examples of such actions include deciding whether to send a packet in a given time slot~\cite{leong2015optimality}, adjusting the transmission power to improve communication quality~\cite{liu2022rollout}, or refraining from using the channel.
For simplicity, we assume the action set to be finite.

We consider a communication channel with a hidden binary mode $\theta_t \in \iTheta = \{\tg, \tb\}$, which evolves according to a Markov process.
The two modes $\tg$ and $\tb$ correspond to favorable and unfavorable channel conditions, respectively, satisfying $\lambda(\tg,a) \ge \lambda(\tb,a)$ for any $a \in \mcal{A}$, where $\lambda(\theta,a)$ denotes the probability of successful transmission under action $a$.
The channel mode evolves according to the transition probability $\Pc(\theta_{t+1} | \theta_t, a_t)$.
This modeling framework captures a broad class of communication environments with temporally correlated uncertainties, including the Gilbert--Elliott (GE) channel~\cite{mushkin1989capacity} and persistent failure models with geometrically distributed change time.


We make the following assumption.
\begin{assum}\label{assum:mode_TP2}
    The channel mode transition kernel $\Pc(\theta'|\theta,a)$ is TP2 in $(\theta,\theta')$ for any $a\in\mcal{A}$.
\end{assum}
Assumption~\ref{assum:mode_TP2} is satisfied by many commonly used channel models.
For instance, in the GE channel model, it is typically assumed that $\Pc(0|0,a)\gg\Pc(1|0,a)$, and similarly for the channel mode $\theta=1$.
This dominance of self-transition probabilities yields a positive cross-difference term, and hence it has the TP2 property.
In the persistent failure model, once the channel mode switches from $0$ to $1$, it remains in that mode thereafter, where $\Pc(0|1,a)=0$ and it implies a positive cross-difference term.

\subsection{POMDP Formulation}

We formulate the transmission scheduling problem as a POMDP.
The system state is given by the pair $(\tau_t,\theta_t)$.
Accordingly, the state space is defined as $\mcal{X}\coloneqq \mcal{T}\times\iTheta$, the action space as $\mcal{A}$, and the observation space as $\mcal{Y}\coloneqq\mcal{T}$.
The state transition kernel factorizes as
$P(\tau',\theta'|\tau,\theta,a) = \Ph(\tau'|\tau,\theta',a)\Pc(\theta'|\theta,a),$
where the holding-time transition kernel $\Ph$ is given by
\[
\Ph(\tau'|\tau,\theta',a) \coloneqq
 \left\{
 \begin{array}{ll}
  \lambda(\theta',a), & {\rm if}\ \tau'=0,\\
  1-\lambda(\theta',a), & {\rm if}\ \tau'=\tau+1,\\
  0, & {\rm otherwise}.
 \end{array}
 \right.
\]
The initial distributions are given by $\tau_0\sim \pt$ and $\theta_0\sim \pc$, respectively.
Since the holding time is perfectly observed through ACK, the observation kernel is deterministic and given by $B(y|\tau,\theta)=\mathbb{I}\{y=\tau\}$, where $\mathbb{I}$ denotes the indicator function.
The instantaneous cost is defined as
$c(\tau,\theta,a)\coloneqq c_{\rm s}(\tau)+c_{\rm a}(a),$
which captures the trade-off between estimation performance and the cost associated with the scheduling action.
We consider an infinite-horizon discounted cost criterion with discount factor $\gamma\in(0,1)$.

To facilitate analysis, we consider the equivalent belief MDP formulation.
Let $b_t\in[0,1]$ denote the belief that the channel is in the unfavorable state.
We denote the probability mass function of the belief by $\beta(\theta)$, i.e., $\beta(\tg)=1-b$ and $\beta(\tb)=b$.
The initial belief is $b_0=\pc(\tb)$.
Given the current state $(\tau_t,b_t)$, action $a_t$, and observation $y_t=\tau_{t+1}$, the belief is updated by Bayes' rule
$b_{t+1}=T(\tau_t,b_t,y_t,a_t)$
where
\[
\begin{array}{ll}
 T(\tau,b,y,a) \hs &
 = K(\tau,\tb,y,a)\hat{\beta}(\tb;b,a)/\sigma(\tau,b,y,a)
\end{array}
\]
with the predictive belief, the composite kernel, and the observation likelihood
\[
\left\{
\begin{array}{ll}
 \hat{\beta}(\theta;b,a)\hs &\coloneqq \sum_{\phi\in\iTheta}\Pc(\theta|\phi,a)\beta(\phi),\\
 K(\tau,\theta,y,a)\hs &\coloneqq \sum_{\tau'\in\mcal{T}}B(y|\tau',\theta)\Ph(\tau'|\tau,\theta,a),\\
 \sigma(\tau,b,y,a)\hs & \coloneqq \sum_{\theta\in\iTheta} K(\tau,\theta,y,a)\hat{\beta}(\theta;b,a),
\end{array}
\right.
\]
respectively.
The belief MDP has the augmented state space $\mcal{S}_{\rm b}\coloneqq\mcal{T}\times[0,1]$.
The expected instantaneous cost under belief $b$ is given by
$c(\tau,b,a)\coloneqq c(\tau,\tg,a)(1-b)+c(\tau,\tb,a)b=c_{\rm s}(\tau)+c_{\rm a}(a)$.
The belief MDP is thus described by the tuple $(\mcal{S}_{\rm b},\mcal{A},\mcal{Y},P,B,c,\gamma,(\pt,b_0))$.
Our objective is to find an optimal belief-dependent policy $\pi(a|\tau,b)$ that minimizes the infinite-horizon discounted cumulative cost
$J = \Exp_\pi\left[\sum_{t=0}^\infty \gamma^t c(\tau_t,b_t,a_t)\right].$

A subtle technical issue arises from the fact that the state space is countably infinite and the cost function becomes unbounded when the system matrix $\rmA$ is not Schur stable, i.e., $\rho(\rmA)>1$.
To keep the focus on our main objective, we impose the following assumption, which is standard in the literature to guarantee the convergence of the estimation error in the fully observable setting (e.g.,~\cite[Assumption~3]{wei2023double}).
\begin{assum}\label{assum:success_trans_prob}
    The successful transmission probability satisfies
    $\ul{\lambda}\coloneqq\min_{\theta\in\iTheta,a\in\mcal{A}}\lambda(\theta,a)>1-1/\rho(\rmA)^2$.
\end{assum}
Under Assumption~\ref{assum:success_trans_prob}, we can derive the corresponding Bellman equation in the partially observable setting.
\begin{prop}\label{prop:Bellman_equation}
    Let Assumption~\ref{assum:success_trans_prob} hold.
    Then there exists a $Q$-function that satisfies the Bellman equation
    $\mathfrak{T}Q=Q$ where $\mathfrak{T}$ denotes the Bellman operator defined by
    \[
    \begin{array}{l}
     (\mathfrak{T}Q)(\tau,b,a)\coloneqq c(\tau,b,a)\\
     \quad+\gamma\sum_{y\in\mcal{Y}}\min_{a'\in\mcal{A}}Q(y,T(\tau,b,y,a),a')\sigma(\tau,b,y,a).
    \end{array}
    \]
    Accordingly, the value function and the optimal policy are given by $V(\tau,b)= \min_{a\in\mcal{A}}Q(\tau,b,a)$ and $\pi(\tau,b)=\argmin_{a\in\mcal{A}}Q(\tau,b,a),$ respectively.
\end{prop}
The proof is provided in Appendix.

\section{MONOTONICITY PROPERTIES}
\label{sec:monotonicity_analysis}

Existing studies have revealed that, under fully observable settings, the value function exhibits monotonicity properties with respect to both the holding time and the channel mode~\cite{leong2015optimality,qi2017optimal,wu2018optimal,wu2020learning,wei2023double,sun2024optimal}.
The underlying intuition is that the system performance degrades as the holding time becomes longer or as the channel mode deteriorates, which consequently worsens the value function.
Such monotonicity properties directly imply a threshold structure of the optimal policy, whereby the optimal action switches when the system state crosses certain critical values.
This structural result not only provides valuable insight into the nature of the optimal policy but also significantly reduces the computational complexity of policy computation and implementation~\cite{roy2021online,nakhleh2022deeptop}.
The objective of this section is to extend this line of analysis to our partially observable setting and investigate whether similar monotonicity properties can be established for the belief-state value function.

\subsection{Issue: Lack of Order Preservation in Transition Kernel}

A common approach to monotonicity analysis for POMDPs is to exploit the TP2 properties of the transition kernels.
It is well known that, if the state transition kernel $P$ and the observation kernel $B$ are TP2 for all actions, then the value function inherits a monotonicity property, provided that the instantaneous cost function is monotone (see Prop.~\ref{prop:monotonicity_cost} for our setting)~\cite[Sec.~12.2]{krishnamurthy2025partially}.
Moreover, since the product of TP2 kernels remains TP2, it suffices to examine the TP2 properties of the individual kernels $\Ph,\Pc,B$.

However, as pointed out in~\cite{liu2022rollout,sun2025optimal}, the holding-time transition kernel $\Ph$ is \emph{never} TP2 in $(\tau,\tau')$.
This fact can be readily verified by a simple counterexample.
Specifically, consider $(\tau_1,\tau'_1)=(0,0)$ and $(\tau_2,\tau'_2)=(2,1)$.
Then
\begin{equation}\label{eq:cross_difference}
 \Ph(\tau'_1|\tau_1)\Ph(\tau'_2|\tau_2) - \Ph(\tau'_2|\tau_1)\Ph(\tau'_1|\tau_2) = -(1-\lambda)\lambda<0
\end{equation}
where the dependence on $\theta$ and $a$ is omitted for notational simplicity.
Since TP2 requires this cross-difference to be nonnegative, the kernel $\Ph$ fails to satisfy the TP2 property.
As a consequence, standard TP2-based monotonicity arguments do not directly apply, which has hindered the investigation of monotonicity properties in partially observable settings involving holding-time dynamics.

\subsection{Core Idea: State-Space Folding}

\begin{figure}[t]
  \centering
  \includegraphics[width=0.98\linewidth]{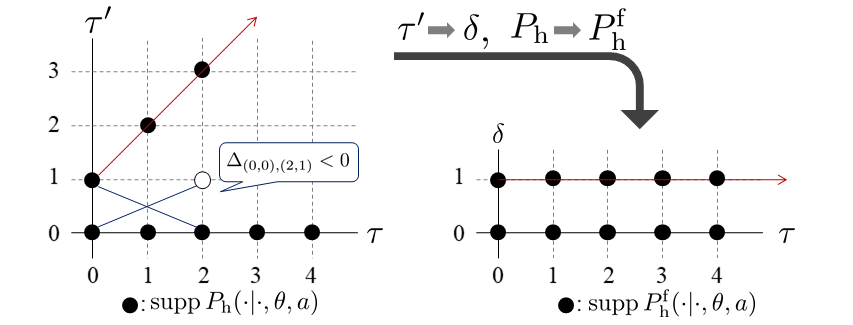}
  \caption{
  State-space folding.
  }
  \label{fig:state_space_folding}
\end{figure}

The key observation is that this non-TP2 behavior is not intrinsic to the transition kernel itself, but stems from the structure of ${\rm supp}\,\Ph(\cdot|\cdot,\theta,a)$, as illustrated in the left part of Fig.~\ref{fig:state_space_folding}, where $\Delta_{(0,0),(2,1)}$ denotes the cross-difference in~\eqref{eq:cross_difference}.

Our core idea, termed \emph{state-space folding}, is to fold the space of the subsequent state so as to collapse states outside ${\rm supp}\,\Ph(\cdot|\cdot,\theta,a)$ and to conduct the monotonicity analysis on the resulting folded space.
Specifically, we introduce a binary set $\mcal{D}\coloneqq\{0,1\}$, representing transmission success and failure, respectively, and define the transition kernels on the folded space as
\[
\begin{array}{l}
 \Pfh(\delta|\tau,\theta,a)  \coloneqq \left\{
 \begin{array}{ll}
 \Ph(0|\tau,\theta,a), & {\rm if}\ \delta=0,\\
 \Ph(\tau+1|\tau,\theta,a), & {\rm if}\ \delta=1,
 \end{array}
 \right.\\
 \Bf(\zeta|\delta,\theta) \coloneqq \left\{
 \begin{array}{ll}
 1, & {\rm if}\ \delta=\zeta,\\
 0, & {\rm otherwise}.
 \end{array}
 \right.
\end{array}
\]
This construction yields the composite kernel
\[
 \textstyle{
 \Kf(\tau,\theta,\zeta,a) \coloneqq \sum_{\delta\in\mcal{D}}\Bf(\zeta|\delta,\theta)\Pfh(\delta|\tau,\theta,a).
 }
\]
The right part of Fig.~\ref{fig:state_space_folding} illustrates the folding operation that removes states outside the support.
Note that the original observed signal is recovered by $y=0$ if $\zeta=0$ and $y=\tau+1$ if $\zeta=1$.
We denote this value by $y(\tau,\zeta)$ with abuse of notation.

This transformation is well-defined, in the sense that they are equivalent under the transformation of the observed signal.
\begin{prop}\label{prop:K_identical}
    The kernels satisfy $K(\tau,\theta,y(\tau,\zeta),a)=\Kf(\tau,\theta,\zeta,a)$.
\end{prop}
\begin{proof}
    It is straightforward from the definition.
\end{proof}
This transformed kernel induces the Bayesian update and the observation likelihood over the folded space given by
\[
\begin{array}{ll}
 \Tf(\tau,b,\zeta,a) \hs & \coloneqq \Kf(\tau,1,\zeta,a)\hat{\beta}(1;b,a)/\sigmaf(\tau,b,\zeta,a),\\
 \sigmaf(\tau,b,\zeta,a) \hs & \coloneqq \sum_{\theta\in\iTheta} \Kf(\tau,\theta,\zeta,a)\hat{\beta}(\theta;b,a).
\end{array}
\]
The bellman operator can be rewritten in terms of them by
\begin{equation}\label{eq:Bellman_folded}
\begin{array}{l}
 (\mathfrak{T}Q)(\tau,b,a) = c(\tau,b,a)\\
 \quad \displaystyle{
 +\gamma\sum_{\zeta\in\mcal{D}}\min_{a'\in\mcal{A}}Q(y(\tau,\zeta),\Tf(\tau,b,\zeta,a),a')\sigmaf(\tau,b,\zeta,a).
 }
\end{array}
\end{equation}

It should be noted that this folding operation does not transform the state space itself; rather, it only affects the subsequent state space and the observation space.
Therefore, this is not a transformation of the POMDP itself, but merely a modification of the transition kernels.

\subsection{Monotonicity Properties}

We examine the TP2 properties of the transformed kernels.
\begin{lem}\label{lem:kernels_TP2}
    The kernel $\Pfh(\delta|\tau,\theta,a)$ is TP2 in $(\delta,\tau)$ and $(\delta,\theta)$.
    The kernel $\Bf(\zeta|\delta,\theta)$ is TP2 in $(\zeta,\delta)$.
    Further, the kernel $\Kf(\tau,\theta,\zeta,a)$ is TP2 in $(\tau,\zeta)$ and $(\theta,\zeta)$.
\end{lem}
\begin{proof}
    For $(\delta,\tau)$ in $\Pfh,$ consider the cross-difference $\Delta_{\tau,\delta}\coloneqq\Pfh(0|\tau_1,\theta,a)\Pfh(1|\tau_2,\theta,a)-\Pfh(1|\tau_2,\theta,a)\Pfh(0|\tau_1,\theta,a)$ for $\tau_1\leq\tau_2$.
    It can be confirmed that $\Delta_{\tau,\delta}=0$ because $\Pfh$ is independent of $\tau$.
    For $(\delta,\theta),$ consider $\Delta_{\theta,\delta}\coloneqq\Pfh(0|\tau,\tg,a)\Pfh(1|\tau,\tb,a)-\Pfh(1|\tau,\tg,a)\Pfh(0|\tau,\tb,a)=\lambda(\tg,a)(1-\lambda(\tb,a))-(1-\lambda(\tg,a))\lambda(\tb,a)=\lambda(\tg,a)-\lambda(\tb,a)\geq 0$ from the property of the channel modes.
    The TP2 property of $\Bf$ in $(\zeta,\delta)$ is straightforward from the definition.
    For $(\tau,\zeta)$ in $\Kf$, Lemma~\ref{lem:TP2_preservation} leads to the TP2 property since $\Kf$ is defined by marginalization with respect to $\delta$.
    Finally, because $\Bf(\zeta|\delta,\theta)$ is independent of $\theta$ and $\Kf$ is defined by marginalization with respect to $\delta$, $\Kf$ is TP2 in $(\theta,\zeta)$.
\end{proof}

These properties, in turn, yield monotonicity of the belief update and the observation likelihood.
\begin{lem}\label{lem:monotonicity_T_sigma}
    Let Assumption~\ref{assum:mode_TP2} hold.
    The Bayesian update over the folded space $\Tf(\tau,b,\zeta,a)$ is increasing in $(\tau,b,\zeta)$.
    The observation likelihood over the folded space satisfies $\sigmaf(\tau_1,b_1,\cdot,a)\leq_{\rm r}\sigmaf(\tau_2,b_2,\cdot,a)$ for $\tau_1\leq \tau_2$ and $b_1\leq b_2$.
\end{lem}
\begin{proof}
    First, since $\Tf$ is indepdendent of $\tau$, it is increasing in $\tau$.
    From Assumption~\ref{assum:mode_TP2} and Lemma~\ref{lem:TP2_transformation}, $\hat{\beta}(1;b,a)$ is increasing in $b$.
    Since $\Kf(\tau,\theta,\zeta,a)$ is TP2 in $(\theta,\zeta)$ from Lemma~\ref{lem:kernels_TP2}, Lemma~\ref{lem:Bayes_order_preservation} implies that $\Tf(\tau,b,\zeta,a)$ is increasing in $b$.
    Further, Lemma~\ref{lem:Bayes_order_preservation} implies that $\Tf(\tau,b,\zeta,a)$ is increasing in $\zeta$.

    Next, since $\hat{\beta}(\theta;b,a)\geq0$ is independent of $\tau,\zeta$ and Lemma~\ref{lem:kernels_TP2}, Lemma~\ref{lem:TP2_weighted_sum} implies that $\sigmaf$ is TP2 in $(\tau,\zeta)$.
    Further, since $\sigmaf$ is given by marginalization with respect to $\theta$, Lemmas~\ref{lem:TP2_preservation} and~\ref{lem:kernels_TP2} imply that $\sigmaf$ is TP2 in $(b,\zeta)$.
    Those are equivalent to the MLR dominance.
\end{proof}

These results lead to our main objective: monotonicity of the value function.
\begin{theorem}\label{thm:monotonicity}
    Under Assumptions~\ref{assum:mode_TP2} and~\ref{assum:success_trans_prob}, the value function $V(\tau,b)$ is increasing in $\tau$ and $b$.
    Furthermore, the $Q$-function $Q(\tau,b,a)$ is also increasing in $\tau$ and $b$ for any $a$.
\end{theorem}
\begin{proof}
    Define $Q_{n+1}\coloneqq \mathfrak{T} Q_n$ with $Q_0=0$ inductively.
    We show the claim by induction.
    Assume that $Q_n(\tau,b,a)$ is increasing in $\tau$ and $b$.
    Since $\Tf(\tau,b,\zeta,a)$ is increasing in $(\tau,b,\zeta)$ from Lemma~\ref{lem:monotonicity_T_sigma} and $y(\tau,\zeta)$ is increasing in $(\tau,\zeta)$, $Q_n(y(\tau,\zeta),\Tf(\tau,b,\zeta,a),a')$ is increasing in $(\tau,b,\zeta)$.
    Because taking minimum preserves monotonicity, $\min_{a'}Q_n(y(\tau,\zeta),T(\tau,b,\zeta,a),a')$ is increasing in $(\tau,b,\zeta)$.
    From Lemmas~\ref{lem:MLR_dom_FSD} and~\ref{lem:monotonicity_T_sigma}, $\gamma\sum_{\zeta\in\mcal{D}}\min_{a'} Q_n(y(\tau,\zeta),\Tf(\tau,b,\zeta,a),a')\sigmaf(\tau,b,\zeta,a)$ is increasing in $\tau,b$.
    Since summation preserves monotonicity, Prop.~\ref{prop:monotonicity_cost} and~\eqref{eq:Bellman_folded} imply that $Q_{n+1}(\tau,b,a)$ is increasing in $\tau$ and $b$.
    Then $Q$, the limit of $Q_n$, is also increasing in $\tau$ and $b$.
    Therefore, $V(\tau,b)$ is increasing in $\tau$ and $b$ as well.
\end{proof}

\subsection{Structural Threshold Property of Optimal Stopping Policy}

As a representative consequence of the monotonicity property, we examine an associated optimal stopping formulation.
In particular, we consider a binary action space $\mcal{A}=\{0,1\}$, where $0$ and $1$ correspond to continuation and termination, respectively.
Once $a=1$ is selected, the process terminates immediately and incurs a stopping cost $c_{\rm a}(1)=c_{\rm stop}\in\mathbb{R}$.
Without loss of generality, we assume $c_{\rm a}(0)=0$.
This formulation naturally captures scenarios such as the quickest detection of a persistent mode change induced by equipment failure or adversarial intrusion.
Under this setting, the Bellman equation is given by
\[
\left\{
\begin{array}{l}
     Q(\tau,b,0)\coloneqq c(\tau,b,0)\\
     \quad+\gamma\sum_{y\in\mcal{Y}}\min_{a'\in\mcal{A}}Q(y,T(\tau,b,y,0),a')\sigma(\tau,b,y,0),\\
    Q(\tau,b,1)\coloneqq c_{\rm stop}.
\end{array}
\right.
\]

The monotonicity directly leads to the structural threshold property of the optimal stopping policy.
\begin{theorem}\label{thm:threshold}
    Let Assumptions~\ref{assum:mode_TP2} and~\ref{assum:success_trans_prob} hold.
    In the optimal stopping setting, there exists a monotone threshold function
    $b_{\rm th}:\mcal{T}\to[0,1]$ such that the optimal policy satisfies
    $\pi(\tau,b)=1$ if $b \geq b_{\rm th}(\tau)$ and $\pi(\tau,b)=0$ otherwise, where $b_{\rm th}(\tau)$ is decreasing in $\tau$.
\end{theorem}
\begin{proof}
    Since $c_{\rm stop}$ is a constant and $-Q(\tau,b,0)$ is decreasing from Theorem~\ref{thm:monotonicity},
    $\Delta_Q(\tau,b)\coloneqq Q(\tau,b,1)-Q(\tau,b,0)=c_{\rm stop}-Q(\tau,b,0)$ is decreasing, and hence the Topkis' theorem (Lemma~\ref{lem:Topkis}) leads to the claim.
\end{proof}

Theorem~\ref{thm:threshold} implies that the optimal stopping policy is characterized by a monotone threshold function $b_{\rm th}(\tau)$.
Intuitively, as the holding time increases or the belief of the unfavorable channel mode becomes stronger, the expected benefit of continuing diminishes, and stopping becomes optimal once it exceeds the critical level.

\section{NUMERICAL EXAMPLE}
\label{sec:num}

We validate the theoretical results through a numerical example in an optimal stopping setting.
The system parameters are given by $\rmA=0.85, \rmC=1.0, \rmQ=0.3, \rmR=0.3$.
The channel parameters are $\lambda(\tg)=0.9,\lambda(\tb)=0.2,\Pc(\tg|\tg)=0.9,\Pc(\tb|\tb)=1.0$.
The stopping cost is $c_{\rm stop}=10.0$, and the discount factor is $\gamma=0.95$.
The optimal policy is computed via value iteration with pruning~\cite{cassandra1997incremental} implemented using {\tt pomdp\_py}~\cite{zheng2020pomdp_py}.
The code is available online~\cite{Sasahara2026code}.

\begin{figure}[t]
  \centering
  \includegraphics[width=0.97\linewidth]{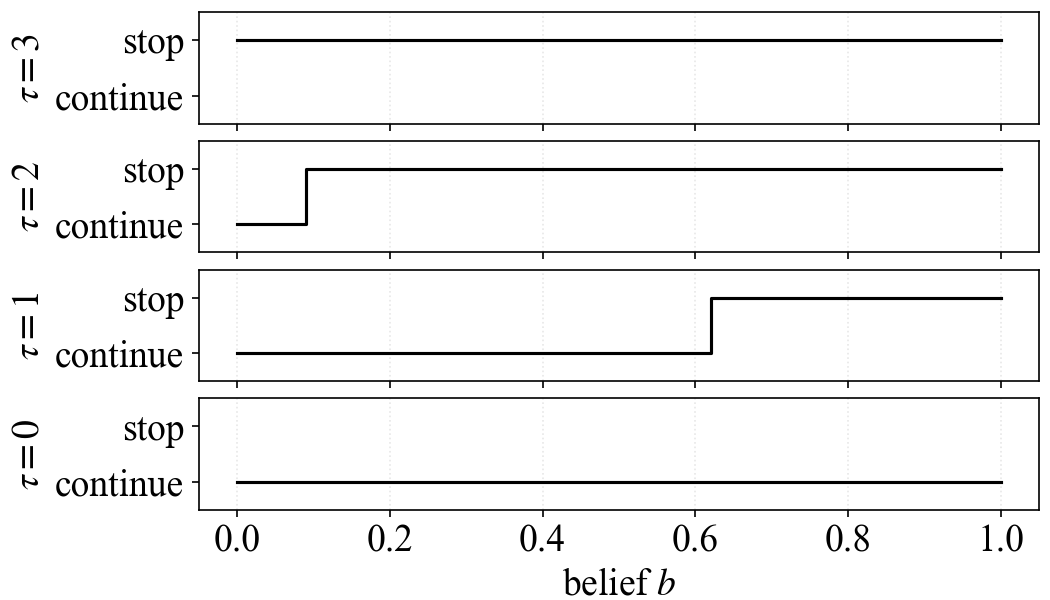}
  \caption{
  Threshold structure of optimal stopping policy.
  }
  \label{fig:threshold_policy}
\end{figure}

Fig.~\ref{fig:threshold_policy} illustrates the optimal policy.
Each subfigure is associated with a specific holding time $\tau$, where the horizontal axis represents the belief and the vertical axis represents the corresponding optimal action.
We first observe that the optimal policy exhibits a threshold structure, where the stop action is taken when the belief exceeds a certain value that depends on the holding time.
We further observe that the threshold $b_{\rm th}(\tau)$ is decreasing in $\tau$.
This result is consistent with Theorem~\ref{thm:threshold}.

\section{CONCLUSION}
\label{sec:conc}

In this paper, we have examined transmission scheduling for remote state estimation with hidden binary modes by formulating the problem as a POMDP and introducing a state-space folding technique to recover order-preserving properties of the transformed transition kernels.
Leveraging TP2-based analysis on the folded space, we have established monotonicity in both the holding time and the belief of the unfavorable channel mode.
As a representative application, we have proved that the optimal stopping policy admits a monotone threshold structure.
Future work includes the development of computationally efficient algorithms for optimal policy design, extensions to more general scenarios, and game-theoretic analyses of security settings in which the channel mode is manipulated by an adversary.








\appendix

We prove Prop.~\ref{prop:Bellman_equation} beginning with the following lemma.
\begin{lem}\label{lem:Gelfand_approx}
    For any $\epsilon>0$, there exists $C>0$ such that $\|\rmA^\tau\|_2\leq C(\rho(\rmA)+\epsilon)^\tau$ for any $\tau\in\mcal{T}$.
\end{lem}
\begin{proof}
    First, from the Gelfand's formula~\cite{kozyakin2009accuracy}, we have $\lim_{\tau\to\infty}\|\rmA^\tau\|_2^{1/\tau}=\rho(\rmA)$.
    Thus, for any $\epsilon>0$ there exists $\ul{\tau}\in\mcal{T}$ such that $\|\rmA^\tau\|_2\leq(\rho(\rmA)+\epsilon)^\tau$ for any $\tau>\ul{\tau}$.
    Hence, for sufficiently large $\tau,$ the desired inequality holds with $C=1$.
    Since the set of $\tau$ for which the inequality does not hold with $C=1$ is finite,
    there exists a sufficiently large constant $C>0$ such that it holds for any $\tau\in\mcal{T}$.
\end{proof}

Define the weighted sup-norm of a function $f:\mcal{T}\times[0,1]\times\mcal{A}\to\mathbb{R}$ as $\sup_{\tau,b,a}|f(\tau,b,a)|/s(\tau)$ with a positive-valued function $s(\tau)\coloneqq (\rho(\rmA)+\epsilon)^{2\tau}$.
Define also the function space $\mcal{B}$ as the set of all functions being bounded with respect to the norm.
The following lemma holds.
\begin{lem}\label{lem:convergence_properties}
    The three properties hold:
    (a) $c(\tau,b,a)\in\mcal{B}$,
    (b) $\sum_{y\in\mcal{T}}\sigma(\tau,b,y,a)s(y)\in\mcal{B}$,
    (c) Under Assumption~\ref{assum:success_trans_prob}, for any $\epsilon>0$ that satisfies $\alpha\coloneqq(1-\ul{\lambda})(\rho(\rmA)^2+\epsilon)<1$ there exists $m\in\mathbb{N}$ such that $\gamma^m \sum_{y\in\mcal{Y}}\Prob(\tau_m=y|\tau_0=\tau)s(y)/s(\tau)<1$ for any $\tau\in\mcal{T}.$
\end{lem}
\begin{proof}
    For (a), take $q>0$ such that $\ol{\rmP}\preceq qI$ and $\rmQ\preceq qI$.
    From the commutative property of trace, $\tr{XY}\leq\rho(X)\tr{Y}$, and $\rho(X^{\sf T}X)=\|X\|_2^2$, Lemma~\ref{lem:Gelfand_approx} implies that
    \[
     \begin{array}{l}
     c_{\rm s}(\tau) = \tr{(\rmA^{\sf T}\rmA)^\tau \ol{\rmP}}+\sum_{t=0}^{\tau-1}\tr{(\rmA^{\sf T}\rmA)^t\rmQ}\\
     \leq \tr{(\rmA^{\sf T}\rmA)^\tau} \tr{\ol{\rmP}} + \sum_{t=0}^{\tau-1}\tr{(\rmA^{\sf T}\rmA)^t}\tr{\rmQ}\\
     \leq q \sum_{t=0}^\tau \tr{(\rmA^{\sf T}\rmA)^t}\leq q \sum_{t=0}^\tau \|\rmA^t\|_2^2\leq qC^2s(\tau).
     \end{array}
    \]
    Since
    $c_{\rm a}(a)$ is bounded, the property (a) holds.

    For (b), note that the inequality $\sigma(\tau,b,y,a)\leq 1$ trivially holds.
    Thus $\sum_{y\in\mcal{Y}}\sigma(\tau,b,y,a)s(y)/s(\tau)=(\sigma(\tau,b,0,a)s(0)+\sigma(\tau,b,\tau+1,a)s(\tau+1))/s(\tau)\leq (s(0)+s(\tau+1))/s(\tau)\leq 1/(\rho(\rmA)+\epsilon)^{2\tau} + (\rho(\rmA)+\epsilon)^2<+\infty$.

    Finally, we show (c).
    For $y = 0,\ldots,m-1$, the event $\tau_m = y$ means that a successful transmission occurs at time $m-y$, followed by $y$ consecutive transmission failures.
    Hence, $\Prob(\tau_m=y|\tau_0=\tau)\leq(1-\ul{\lambda})^y$.
    Similarly, $\Prob(\tau_m=\tau+m|\tau_0=\tau)\leq(1-\ul{\lambda})^m$.
    Further, other probabilities are zero.
    Thus,
    $\sum_{y\in\mcal{Y}} \Prob(\tau_m=y|\tau_0=\tau)s(y) \leq \sum_{y=0}^{m-1} \alpha ^y + \alpha^m s(\tau)$.
    From the hypothesis in the claim,
    $\gamma^m\sum_{y=0}^{m-1} \alpha^y/s(\tau) \leq \gamma^m/\{(1- \alpha)(\rho(\rmA)+\epsilon)^{2\tau}\}\to 0$
    as $m\to \infty.$
    In addition, since $\gamma^m\alpha^m s(\tau)/s(\tau)=\gamma^m\alpha^m\to0$ as $m\to \infty$, the desired inequality holds for sufficiently large $m\in\mathbb{N}$.
\end{proof}

Based on Lemma~\ref{lem:convergence_properties}, we can show that the $m$-fold composition of the Bellman operator
is a contraction map over $\mcal{B}$, by closely following the existing proof~\cite[Sec.~1.5]{bertsekas2012dynamicII}.
Finally, Prop.~\ref{prop:Bellman_equation} is straightforward from the contraction property.

\begin{proof}
    Since $\mathfrak{T}^m$ is a contraction map over $\mcal{B}$ for sufficiently small $\epsilon>0$, the $m$-stage contraction mapping fixed-point theorem~\cite[Prop.~1.5.4]{bertsekas2012dynamicII} leads to the claim.
\end{proof}


\bibliographystyle{IEEEtran}
\bibliography{CDC2026refs}

\end{document}